\documentclass{article}

\usepackage{microtype}
\usepackage{graphicx}
\usepackage{subfigure}
\usepackage{booktabs} % for professional tables
\usepackage{enumitem}

\usepackage{hyperref}

\usepackage[accepted]{icml2025}

\usepackage{amsmath}
\usepackage{amssymb}
\usepackage{mathtools}
\usepackage{amsthm}

\usepackage[capitalize,noabbrev]{cleveref}

\theoremstyle{plain}

\theoremstyle{definition}

\theoremstyle{remark}

\usepackage[textsize=tiny]{todonotes}
\definecolor{mygray}{gray}{.9}

\icmltitlerunning{BinauralFlow: A Causal and Streamable Approach for High-Quality Binaural Speech Synthesis with Flow Matching Models}

\begin{document}

\twocolumn[
\icmltitle{BinauralFlow: A Causal and Streamable Approach for High-Quality Binaural Speech Synthesis with Flow Matching Models}



\icmlsetsymbol{equal}{*}

\begin{icmlauthorlist}
\icmlauthor{Susan Liang}{equal,ur,comp}
\icmlauthor{Dejan Markovic}{comp}
\icmlauthor{Israel D. Gebru}{comp}
\icmlauthor{Steven Krenn}{comp}
\icmlauthor{Todd Keebler}{comp}
\icmlauthor{Jacob Sandakly}{comp}
\icmlauthor{Frank Yu}{comp}
\icmlauthor{Samuel Hassel}{comp}
\icmlauthor{Chenliang Xu}{ur}
\icmlauthor{Alexander Richard}{comp}
\end{icmlauthorlist}

\icmlaffiliation{ur}{University of Rochester, NY, USA}
\icmlaffiliation{comp}{Codec Avatars Lab, Meta, PA, USA}

\icmlcorrespondingauthor{Susan Liang}{sliang22@ur.rochester.edu}
\icmlcorrespondingauthor{Alexander Richard}{richardalex@meta.com}

\icmlkeywords{Machine Learning, ICML}

\vskip 0.3in
]



\printAffiliationsAndNotice{* Work done during an internship at Meta.} % otherwise use the standard text.

\begin{abstract}
Binaural rendering aims to synthesize binaural audio that mimics natural hearing based on a mono audio and the locations of the speaker and listener. 
Although many methods have been proposed to solve this problem, they struggle with rendering quality and streamable inference.
Synthesizing high-quality binaural audio that is indistinguishable from real-world recordings requires precise modeling of binaural cues, room reverb, and ambient sounds. Additionally, real-world applications demand streaming inference.
To address these challenges, we propose a flow matching based streaming binaural speech synthesis framework called BinauralFlow. We consider binaural rendering to be a generation problem rather than a regression problem and design a conditional flow matching model to render high-quality audio. Moreover, we design a causal U-Net architecture that estimates the current audio frame solely based on past information to tailor generative models for streaming inference. Finally, we introduce a continuous inference pipeline incorporating streaming STFT/ISTFT operations, a buffer bank, a midpoint solver, and an early skip schedule to improve rendering continuity and speed.
Quantitative and qualitative evaluations demonstrate the superiority of our method over SOTA approaches. A perceptual study further reveals that our model is nearly indistinguishable from real-world recordings, with a $42\%$ confusion rate.
We recommend that readers visit our project page for demo videos: \url{https://liangsusan-git.github.io/project/binauralflow/}.
\end{abstract}
\section{Introduction}
Unlike monaural audio, which conveys content in a single channel with no spatial context, spatial audio presents the audience with a multi-dimensional listening experience by rendering sounds from various directions and distances. 
When rendered using two audio channels and played back to the user's ears through headphones, spatial audio is also referred to as binaural audio. 
Its ability to enhance realism and user engagement makes spatial audio a key component of a wide range of immersive applications, from
cinematic experiences and gaming \cite{raghuvanshi2018parametric,chaitanya2020directional,broderick2018importance,yadegari2024spatial} to rapidly evolving fields such as virtual (VR), augmented (AR) and mixed realities (MR) \cite{zotkin2004rendering,kim2019immersive,gupta2022augmented,schutze2018new,cohen2015special,yang2020effects,kailas2021design,liang2024language,huang2024modeling}\nocite{zhang2021unicon}.

Although a lot of work has been done in both signal processing and machine learning communities \citep{savioja1999creating,zotkin2004rendering,jianjun2015natural,zhang2017surround,gao20192,richard2021neural,leng2022binauralgrad,liang2023neural}, the current state-of-the-art methods still struggle with achieving both (1) \textbf{high-quality} rendering and (2) \textbf{causal and streamable} inference. In particular,  generating high-fidelity binaural audio that is truly indistinguishable from real-world recordings, has remained an open problem. Given a (virtual) acoustic source and its audio signal, rendering binaural audio that is of such quality to deceive the listener into believing it is truly present in the space requires careful consideration and modeling of binaural cues, room reverb, and ambient noise. 
The poses of the sound source and receiver are key to perception. The distance between them primarily affects the overall audio level, while their relative orientation influences the perceived direction of the sound source (e.g., interaural level and time differences). Meanwhile, the inclusion of reverberation effects and background noise that match the environment is crucial for improving the realism and immersion of the acoustic scene. Existing approaches might not fully consider all of these factors, leading to suboptimal rendering performance, with noticeable differences between recorded (real) and generated (virtual) sounds.

Furthermore, real-world audio rendering applications require not only high-fidelity audio generation but also \textbf{continuous, streaming} inference capability that maintains low latency, which is essential for applications where audio must be generated or processed in real time, such as live voice synthesis, interactive gaming, or augmented reality systems. However, most advanced neural rendering approaches \cite{gao20192,leng2022binauralgrad,van2016wavenet,sgmse} do not support continuous synthesis, due to the non-causal model architectures and inefficient multi-step inference procedures.

To achieve \textbf{high-fidelity rendering} and \textbf{continuous inference}, 
we propose a flow matching-based streaming binaural speech generation framework which we will refer to as BinauralFlow. Predicting reverberation effects and background noise using a \textit{regression} approach is challenging because these features are absent from the input audio signal and they exhibit stochastic behavior. Instead, we consider the binaural rendering problem to be a \textit{generative} task. We design a conditional flow matching model to enhance perceptual realism by rendering realistic acoustic effects and dynamic ambient noise. To augment rendered binaural speech with precise binaural cues, we condition the model on the poses of the sound source and receiver to guide speech rendering.

Existing flow matching models typically do not support continuous inference due to non-causal model architectures and multi-step inference requirements. Popular generative frameworks \cite{ddpm,song2020score,stablediffusion,sgmse} commonly use a non-causal U-Net \cite{unet} composed of convolution and attention blocks as backbones. Non-causal convolution kernels and the globally aware attention calculation mechanism break the time causality during rendering. Therefore, we introduce a causal U-Net architecture by meticulously designing causal 2d convolution blocks so that the prediction of the next audio chunk solely relies on the past chunks. 

Moreover, a causal backbone alone is not sufficient for streaming inference because of the multi-step generation process required by generative models. Starting from an initial noise, generative diffusion and flow matching models rely on an iterative denoising process which takes a few steps to complete the generation process. 
To enable continuous generation, we need to ensure time causality for all inference steps. 
To this end, we construct a continuous inference pipeline consisting of streaming STFT/ISTFT operations, a buffer bank, a midpoint solver, and an early skip schedule. In this way, we enable seamless streaming inference for U-Net-based generative models. 

In summary, our contributions are: 
\begin{itemize}[noitemsep,topsep=0pt,partopsep=0pt]
    \item We design a flow matching-based streaming binaural audio synthesis framework to render high-fidelity and continuous audio based on the mono input.
    \item We introduce a conditional flow matching approach to the binaural speech rendering problem by considering the problem from a generative perspective.
    \item We propose a causal U-Net architecture that estimates vector fields solely based on history information. We present a continuous inference pipeline supporting the streaming inference of generative models.
    \item We demonstrate the effectiveness of our approach, showing that our model outperforms existing SOTA approaches with a high margin. A perceptual study shows that our model is nearly indistinguishable from real-world recordings with a $42\%$ confusion rate.
\end{itemize}
\section{Related Work}
Our work is closely related to digital audio rendering, neural audio rendering, and generative models.
\subsection{Digital Audio Rendering}
Digital audio rendering approaches utilize Digital Signal Processing (DSP) techniques to render audio. These approaches \citep{savioja1999creating,zotkin2004rendering,jianjun2015natural,zhang2017surround,chen2020soundspaces,chen2022soundspaces} estimate binaural audio with a series of linear time-invariant systems, including room impulse response (RIR) \cite{lin2006bayesian,szoke2019building,antonello2017room}, head-related transfer function (HRTF) \cite{begault20003,cheng1999introduction}, and additive ambient noise. Due to the simplified geometrical simulation \cite{valimaki2012fifty,savioja2015overview}, non-personalized HRTFs, and the assumed stationary noise, there is a noticeable quality gap between real recordings and generated sounds.

\subsection{Neural Audio Rendering}
Recently, researchers have resorted to deep neural networks to render spatial audio given the powerful fitting capabilities of neural networks. \citet{gao20192} introduce a vision-guided binauralization network to generate binaural audio conditioned on a video frame. \citet{richard2021neural} design a neural warp network to warp the mono audio according to the time delay and the listener position. \citet{chen2023novel} and \citet{liang2023av} utilize vision information to guide binaural audio prediction at novel poses. Although these methods achieve plausible speech results, their regression mechanism limits their generation capability, i.e., they cannot generate precise room acoustics and ambient noise that are absent from the input data.

\subsection{Generative Models}
Generative models, especially diffusion models \cite{ddpm,ddim,song2020score}, exhibit strong generative capabilities in the audio domain \cite{diffsound,audioldm,make-an-audio,diffwave,leng2022binauralgrad}. Based on DiffWave \cite{diffwave}, \citet{leng2022binauralgrad} propose a two-stage diffusion model (BinauralGrad) to synthesize binaural audio. \citet{sgmse} design a diffusion model for speech enhancement in the complex STFT domain (SGMSE). However, diffusion models require many sampling steps during inference, e.g., $30$ steps. To reduce inference steps while maintaining performance, \citet{flowmatching} introduce flow matching models that simulate the generation process with the optimal transport transformation. Inspired by this, we propose a flow matching-based generative framework that outperforms SGMSE with more efficient inference. We compare our work and other flow matching-based audio models \citep{lee2024periodwave,liu2024rfwave,welker2025flowdec,mehta2024matcha,du2024cosyvoice,lee2024accelerating} in detail in the appendix.
\section{Method}
In this paper, we propose BinauralFlow, a flow matching-based streaming model for binaural speech rendering. We first formulate the task in \cref{sec:task}. To synthesize high-quality binaural audio, we introduce a conditional flow matching model that is conditioned on both pose information and mono input (\cref{sec:flow}). Then, we design a causal U-Net architecture that estimates the current chunk solely relying on the history information (\cref{sec:model}). Finally, we present our continuous inference pipeline that improves rendering continuity and speed in \cref{sec:inference}.

\subsection{Task Definition}
\label{sec:task}
The goal of the binaural rendering task is to synthesize binaural audio (two channels --- one for each listener's ear) $y\in\mathbb{R}^{2 \times N}$, based on the monaural audio (one channel containing speaker's signal) $x\in\mathbb{R}^{N}$, and the poses of the speaker $p_\mathrm{tx}\in\mathbb{R}^{7\times N'}$ and the listener $p_\mathrm{rx}\in\mathbb{R}^{7\times N'}$, where $N$ is the length of an audio clip and $N'$ is the length of a pose sequence. We represent a pose as a combination of position $(\bar{x},\bar{y},\bar{z})\in\mathbb{R}^3$ and quaternion rotation $(\tilde{w},\tilde{x},\tilde{y},\tilde{z})\in\mathbb{R}^4$. To solve this problem, we need to learn a function $f$ that maps the monaural audio to the binaural audio:
\begin{equation}
\label{eq:task}
    y = f(x|p_\mathrm{tx},p_\mathrm{rx}).
\end{equation}
As mentioned in the introduction, learning of this mapping function $f$ is non-trivial because it is required to consider the binaural cues, and include the room reverb and the ambient noise, which usually are not present in the input mono signal and exhibit stochastic behavior. Moreover, $f$ should support continuous rendering in the streaming inference setting. 

\subsection{Conditional Flow Matching Models}
\label{sec:flow}
To address the \textbf{quality} challenge raised by the binaural audio rendering, we design a conditional flow matching model as an instance of the function $f$. We consider the binaural speech rendering problem to be a generative task and use flow matching models to generate binaural sound effects.

Specifically, given an audio pair of the mono audio $x$ and the binaural audio $y$, we first convert them from the time space to the time-frequency space using Short-Time Fourier Transformation (STFT): $\mathbf{x}=\mathrm{STFT}(x)\in\mathbb{C}^{2 \times (\frac{F}{2}+1)\times T}$ and $\mathbf{y}=\mathrm{STFT}(y)\in\mathbb{C}^{2 \times (\frac{F}{2}+1)\times T}$, where $F$ is Discrete Fourier Transform (DFT) length, $T$ is the number of time frames, and $\mathbb{C}$ represents the complex space. We repeat the mono input along the channel dimension to be two-channel so that $\mathbf{x}$ and $\mathbf{y}$ are of the same shape. Then we sample a random noise $\mathbf{z}\sim \mathcal{N}(\mathbf{x}, \sigma^2I)$ which centers around $\mathbf{x}$ with the radius of $\sigma$. To generate the binaural audio $\mathbf{y}$ based on the mono input $\mathbf{x}$, we aim to design a flow that moves from the source data $\mathbf{z}$ to the target data $\mathbf{y}$.

\begin{figure*}[t]
    \centering
    \includegraphics[width=\linewidth]{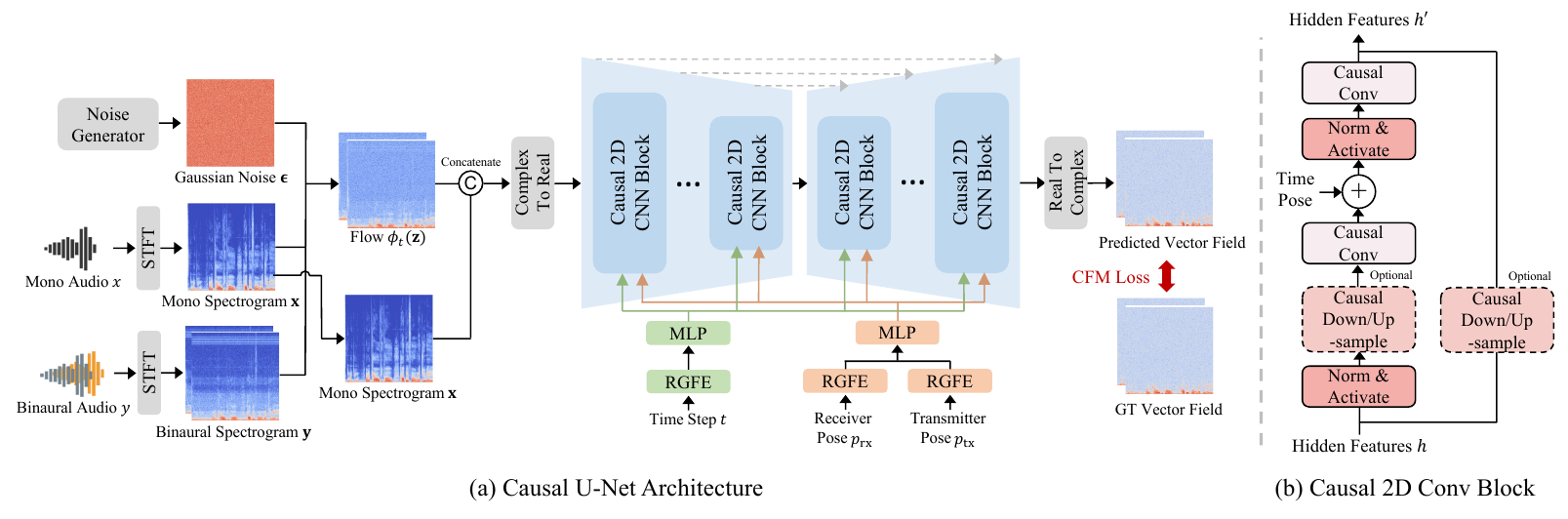}
    \vspace{-6mm}
    \caption{Overview of our BinauralFlow framework. (a) shows the causal U-Net architecture. Our causal U-Net takes as input the flow $\phi_t(\mathbf{z})$ as well as four conditions $t$, $p_\mathrm{rx}$, $p_\mathrm{tx}$, and $\mathbf{x}$, and outputs a predicted vector field. The U-Net consists of several Causal 2D Conv Blocks in the contracting and expanding parts. (b) displays the Causal 2D Conv Block. We design fully causal convolution, down/up-sampling, and normalization layers to ensure temporal causality.}
    \label{fig:framework}
\end{figure*}

We formulate the flow matching problem using the optimal transport formulation inspired by \citet{flowmatching}:
\begin{equation}
\label{eq:flow}
    \phi_t(\mathbf{z}) = t \mathbf{y} + (1-t) \mathbf{z},
\end{equation}
where $\phi_t: [0, 1] \times \mathbb{C}^{2 \times (\frac{F}{2}+1)\times T} \rightarrow \mathbb{C}^{2 \times (\frac{F}{2}+1)\times T}$ is a time-dependent flow function and the flow at time step $t\in[0,1]$ is a linear interpolation between $\mathbf{y}$ and $\mathbf{z}$.

If we use the re-parameterization technique to represent $\mathbf{z}$ as $\mathbf{x}+\sigma \mathbf{\epsilon}$, where $\mathbf{\epsilon}$ is a normal Gaussian noise, $\phi_t$ is updated with
\begin{equation}
\begin{aligned}
\label{eq:reparameter}
    \phi_t(\mathbf{z})&=t\mathbf{y} + (1-t) (\mathbf{x}+\sigma \mathbf{\epsilon}) \\
    &= t\mathbf{y} + (1-t) \mathbf{x} + (1-t)\sigma \mathbf{\epsilon}.
\end{aligned}
\end{equation}
The corresponding probability path $p_t: [0,1] \times \mathbb{C}^{2 \times (\frac{F}{2}+1)\times T} \rightarrow \mathbb{R}_{>0}$ can be calculated as
\begin{equation}
    p_t(\mathbf{z})=\mathcal{N}(\mathbf{z}|t\mathbf{y} + (1-t) \mathbf{x}, (1-t)^2\sigma^2I).
\end{equation}
When $t=0$, $p_0(\mathbf{z})$ is $\mathcal{N}(\mathbf{z}|\mathbf{x}, \sigma^2I)$, which is a Gaussian distribution around the mono audio $\mathbf{x}$ with the radius of $\sigma$. When $t$ gradually increases, the mean of $p_t(\mathbf{z})$ moves linearly from $\mathbf{x}$ to $\mathbf{y}$ and the standard deviation of $p_t(\mathbf{z})$ decreases. If $t=1$, $p_1(\mathbf{z})$ is $\mathcal{N}(\mathbf{z}|\mathbf{y}, 0)$, which collapses to the binaural audio $\mathbf{y}$. Therefore, the flow defined in \cref{eq:flow} moves samples centered around the input audio $\mathbf{x}$ to the binaural audio $\mathbf{y}$ with gradually reduced variance.

Based on the definition of a flow, we can derive a time-dependent vector field $v_t: [0, 1] \times \mathbb{C}^{2 \times (\frac{F}{2}+1)\times T} \rightarrow \mathbb{C}^{2 \times (\frac{F}{2}+1)\times T}$ using the following ordinary differential equation (ODE):
\begin{equation}
    \label{eq:vector}
    \begin{aligned}
        \frac{d}{dt}\phi_t(\mathbf{z}) = v_t(\phi_t(\mathbf{z})).
    \end{aligned}
\end{equation}
By replacing $\phi_t(\mathbf{z})$ in \cref{eq:vector} with \cref{eq:flow}, we calculate the vector field $v_t$ as
\begin{equation}
    v_t(\phi_t(\mathbf{z}))= \mathbf{y} - \mathbf{z}.
\end{equation}
Then we design a deep neural network $u_t$ to match the vector field $v_t$ with the conditional flow matching (CFM) L1 loss:
\begin{equation}
    \label{eq:loss}
    %\mathcal{L}_\mathrm{CFM}=\mathbb{E}_{t, x_0\sim q_0(x), x_1\sim q_1(x), x\sim \mathcal{N}(x_0,\sigma_\mathrm{max})}
    \mathcal{L}_{\mathrm{CFM}}(\theta) = \mathbb{E}_{t, \mathbf{x}, \mathbf{y}, \mathbf{z}} \big | u_t(\phi_t(\mathbf{z}),p_\mathrm{rx},p_\mathrm{tx},\mathbf{x};\theta) - (\mathbf{y} - \mathbf{z}) \big|,
\end{equation}
where $\theta$ is the learnable parameters of the deep neural network $u_t$. We condition the model prediction on the poses of speaker $p_\mathrm{tx}$ and listener $p_\mathrm{rx}$ to accurately model the binaural clues. We also include the mono audio $\mathbf{x}$ to provide rich sound information.

\begin{algorithm}[t]
\caption{Training Procedure of Flow Matching Model}
\label{alg:train}
    \begin{algorithmic}
    \STATE {\bfseries Input:} Dataset $D$, mono audio $\mathbf{x}$, binaural audio $\mathbf{y}$, transmitter pose $p_\mathbf{tx}$, receiver pose $p_\mathbf{rx}$, standard deviation $\sigma$, initial network $u_t$
    \WHILE{not converged}
        \STATE $\{\mathbf{x}, \mathbf{y}, p_\mathbf{tx}, p_\mathbf{rx}\} \sim D$
        // \textcolor{gray}{{\scriptsize Sample data from the dataset}}
        \STATE $\mathbf{z} \sim \mathcal{N}(\mathbf{x},\sigma^2I)$ // \textcolor{gray}{{\scriptsize Sample random variable}}
        \STATE $t \sim \mathcal{U}(0, 1)$ // \textcolor{gray}{{\scriptsize Sample time step}}
        \STATE $\phi_t(\mathbf{z}) \gets t \mathbf{y} + (1 - t) \mathbf{z}$
        %\Comment{Sample a data point on the probability path}
        \STATE $\mathcal{L}_{\mathrm{CFM}}(\theta) \gets | u_t(\phi_t(\mathbf{z}), p_\mathrm{rx}, p_\mathrm{tx}, \mathbf{x};\theta) - (\mathbf{y} - \mathbf{z}) |$
        \STATE $\theta \gets \mathrm{Update}(\theta, \nabla_\theta \mathcal{L}_{\mathrm{CFM}}(\theta))$
    \ENDWHILE
    \STATE {\bfseries Output:} trained network $u_t$
    \end{algorithmic}
\end{algorithm}

We present pseudo code for training a conditional flow matching model in Algorithm \ref{alg:train}. We first select one sample from the dataset. Then we sample a random noise $\mathbf{z}$ following the Gaussian distribution and a time step $t$ following the uniform distribution. We calculate the flow $\phi_t(\mathbf{z})$ at the time step $t$ and pass it along with other conditions into the model $u_t$ to predict the vector field. Finally, we calculate the CFM loss and update the model weights.

\noindent\textbf{Discussion.} Our conditional flow matching method shares some similarity with the simplified flow matching formulation \cite{tong2023improving,jung2024flowavse}. However, we argue that our method is distinct from the simplified flow matching approach. (1) The simplified flow matching approach injects minute perturbation (commonly $1e{-}4$) to the flow, which almost degrades the problem to a deterministic task. Our method uses Gaussian noise of normal magnitude, maintaining the generation randomness. (2) Our method uses mono audio as an important generation condition to improve generation robustness. However, the simplified flow matching model cannot use this condition because it causes model collapse. We provide an experiment in \cref{sec:ablation} to validate the superiority of our approach.

\subsection{Causal U-Net Architecture}
\label{sec:model}
In this section, we describe the proposed network architecture. To tailor the flow matching models for streaming rendering, we design a causal U-Net architecture that predicts the current vector field solely based on past information. 

\begin{figure*}[t]
    \centering
    \includegraphics[width=\linewidth]{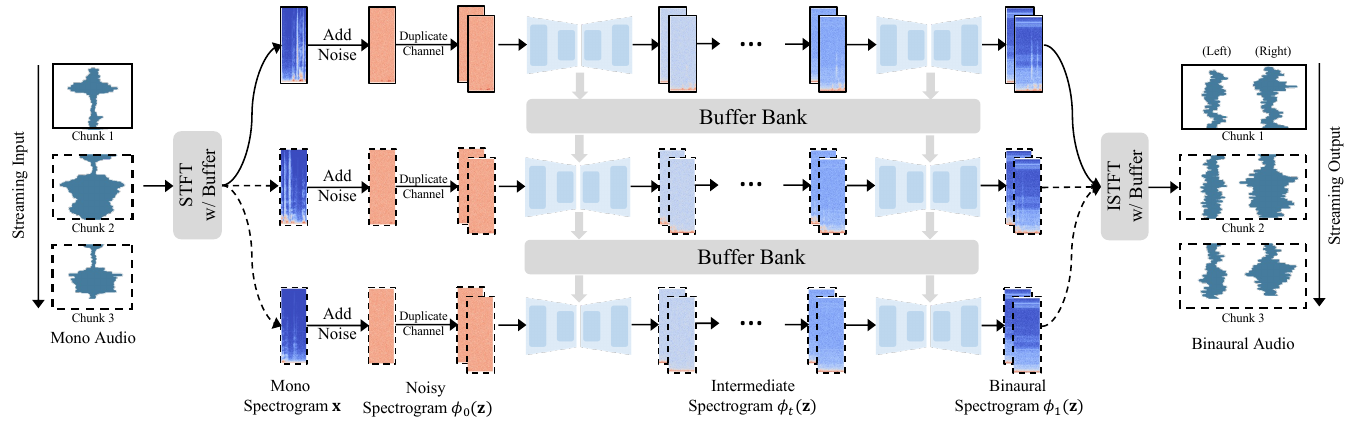}
    \vspace{-5mm}
    \caption{Continuous inference pipeline. Starting with a mono audio chunk (top left, black solid-line box), we compute its spectrogram via streaming STFT, add noise, and duplicate the channel to form the noisy spectrogram $\phi_0(\mathbf{z})$. The trained model progressively removes the noise with a buffer bank. Finally, streaming ISTFT converts the predicted binaural spectrogram $\phi_1(\mathbf{z})$ into binaural audio. When the next audio chunk appears (black dashed-line box), we repeat the process and synthesize seamlessly continuous binaural speech.}
    \label{fig:inference}
\end{figure*}

The complete network architecture is shown in \cref{fig:framework} (a). The input to our network is the flow $\phi_t(\mathbf{z})$ as well as four conditions $t$, $p_\mathrm{rx}$, $p_\mathrm{tx}$, and $\mathbf{x}$, and the output is the predicted vector field. Given a pair of mono and binaural audio signals, $x$ and $y$, we use STFT to calculate their spectrograms $\mathbf{x}$ and $\mathbf{y}$. We sample a normal Gaussian noise $\mathbf{\epsilon}$ of the same shape as $\mathbf{y}$. We compute the flow $\phi_t$ at the time step $t$ using \cref{eq:reparameter}. Then we concatenate $\phi_t$ and the mono spectrogram $\mathbf{x}$ as input to the causal U-Net. Because $\mathbf{x}$ and $\phi_t$ are complex spectrograms, we convert them to real numbers by considering real and imaginary parts as individual channels and concatenating them along the channel dimension. Since both time step $t$ and pose vectors $p_\mathrm{tx}$ and $p_\mathrm{rx}$ are low-dimensional, we employ the positional encoding technique \cite{transformer,nerf} to project them into a high-frequency space. We use Random Gaussian Fourier Embedding (RGFE) \cite{rgfe} followed by Multi-Layer Perceptrons (MLPs) to encode these conditions. The transmitter and receiver pose features are concatenated before feeding into the MLP. We inject the encoded time step and poses into the causal U-Net to guide the vector field prediction. Finally, the causal U-Net estimates the vector field. We convert it back to a complex space using real and imaginary channels. 

Causal U-Net has a contracting part and an expanding part with skip connections between them. Each part consists of several Causal 2D CNN blocks, with architecture shown in \cref{fig:framework} (b). Each block contains Norm and Activate layers, Causal Convolution layers, and optional Causal Down/Up-sampling layers. In the Norm and Activate layer, we utilize GroupNorm \cite{groupnorm} to stabilize training but we limit the computation to each individual frame rather than all frames to ensure causality. We apply the Sigmoid Linear Unit (SiLU) \cite{hendrycks2016gaussian} as an activation function. The Causal Convolution layer is a 3x3 convolution layer with a stride of size 1 and a one-side padding of size 2. One-side padding restricts the receptive field of the convolution kernel to the historical information. Because U-Net requires reducing or increasing the feature dimension in each block, we design a Causal Down/Up-sampling layer. The Causal Downsample layer contains a 4x4 convolution function with a stride of size 2, which reduces the feature dimension by half. The Causal Upsample layer contains a 4x4 transposed convolution function, which doubles the feature dimension. We also add the time step and pose features with the hidden features to guide the vector field prediction. A residual path with an optional Causal Down/Up-sample Layer is included to facilitate learning. 

\subsection{Continuous Inference Pipeline}
\label{sec:inference}
After training BinauralFlow model, we design a continuous inference pipeline to render binaural speech in a streaming manner, as shown in \cref{fig:inference}.
% The continuous inference pipeline is shown in \cref{fig:inference}. 
Given a chunk of mono audio, we apply streaming STFT operation to compute mono spectrogram. We add random noise to it and duplicate its channel to obtain the noisy spectrogram $\phi_0(\mathbf{z})$. Then we use the trained model to gradually remove the injected noise. The denoising process involves several steps and we design a buffer bank to store the buffer of each step. When the next chunk is fed, we retrieve the buffer according to the time step from the buffer bank and reload it to the model. We leverage a midpoint solver and an early skip schedule to improve the denoising speed. Finally, we apply streaming ISTFT to convert the predicted binaural spectrogram $\phi_1(\mathbf{z})$ to binaural audio. When the next audio chunk appears, we repeat the process and generate continuous binaural speech.
Below we describe the individual components that enable continuous inference and improve rendering speed.

\noindent\textbf{Streaming STFT / ISTFT.} We adapt STFT and ISTFT for streaming processing by adding buffers and adjusting the padding manner. We prepend the buffer content to each chunk and update the buffer with the end of the chunk.

\noindent\textbf{Buffer Bank.} In the causal U-Net, we introduce buffers to each causal convolution layer to store the hidden features of the current audio chunk. These buffers are then used to pad the next audio chunk. Since the denoising process involves multiple inference steps, reusing the same buffer across all steps would overwrite historical information. To address this, we construct a dictionary-based buffer bank $B=\{B_t\}_{t=0}^{1}$ to store network buffers of all time steps $t$. During inference, at time step $t$, we retrieve corresponding buffers $B_t$ from the buffer bank. The network buffers $B_t$ are loaded into the U-Net to complete the vector field prediction. Afterword, we store the updated buffers back to the buffer bank to replace $B_t$. We repeat this process until $t$ reaches $1$.

\noindent\textbf{Midpoint Solver.} The inference process requires solving the following ODE to obtain $\phi_1(\mathbf{z})$:
\begin{equation}
    \label{{eq:inference}}
    \begin{aligned}
        \frac{d}{dt}\phi_t(\mathbf{z}) &= u_t(\phi_t(\mathbf{z});\theta), \\
        \phi_0(\mathbf{z})&=\mathbf{z}, \\
    \end{aligned}
\end{equation}
where we omit other model inputs for simplicity. Among different numeric solvers, we choose the Midpoint solver because it effectively reduces the number of function evaluations while maintaining the performance \cite{flowmatching}. We present pseudo code of utilizing the Midpoint solver to solve the ODE in the appendix.

\begin{figure}[t]
    \centering
    \includegraphics[width=\linewidth]{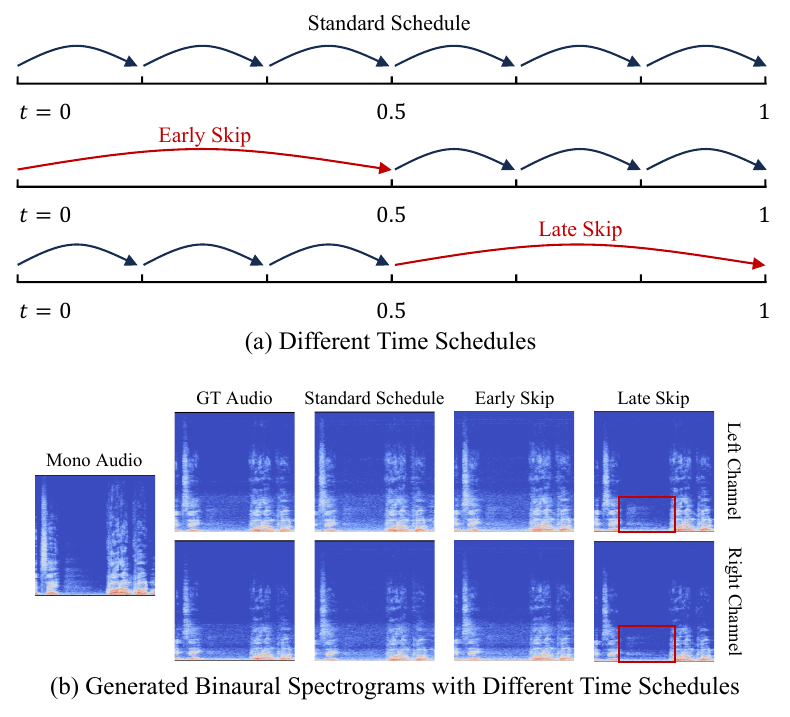}
    \vspace{-6mm}
    \caption{The early skip time schedule. The use of an early skip strategy effectively reduces the inference steps and retains the generation performance.}
    \label{fig:schedule}
    \vspace{-2mm}
\end{figure}

\noindent\textbf{Early Skip Schedule.} To further reduce the number of function evaluations, we propose an early skip schedule. A standard time schedule divides the interval from 0 to 1 into equal segments and moves sequentially from 0 to 1. As shown in \cref{fig:schedule} (a), we design two new schedules: an early skip schedule that skips the first half segments and a late skip schedule that avoids the second half segments. We empirically observe that the use of the early skip schedule does not compromise rendering quality while the late skip degrades the performance, with worse modeling of the background noise (see \cref{fig:schedule} (b)). We speculate that flow matching may be able to correct the errors from the first half during the second half of inference, so even if we conduct early skipping, it does not noticeably affect performance. Therefore, we utilize the early skip strategy to reduce the inference steps to $6$. In comparison, SGMSE model \cite{sgmse} generates comparable results with $30$ steps.

\section{Experiments}
\begin{table*}[t]
\centering
\caption{Quantitative comparison with existing baselines. We show the model type (R: Regression and G: Generation), the number of function evaluations (NFE), the inference speed, and the model size of each approach. The L2 error is on the scale of $1e{-}5$.}
\label{tab:sota_speakeasy}
\resizebox{\linewidth}{!}{
\begin{tabular}{l|c|c|cc|ccc}
\toprule
Methods & Type & NFE & Speed (ms) & Model Size (MB) & L2 $\downarrow$ & Mag $\downarrow$ & Phase $\downarrow$ \\
\midrule
SoundSpaces 2.0 \cite{chen2022soundspaces} & - & 1 & - & - & 4.91 & 0.0129 & 1.58\\
2.5D Visual Sound \cite{gao20192} & R & 1 & 1.1	& 82.0 & 2.78 & 0.0174 & 1.56\\
WaveNet \cite{van2016wavenet} & R & 1 & 21.0 &	32.7 & 2.79 & 0.0175 & 1.57\\
WarpNet \cite{richard2021neural} & R & 1 & 21.9	& 32.8 & 2.79 & 0.0176 & 1.57 \\
\midrule
BinauralGrad \cite{leng2022binauralgrad} & G & 6 & 221.1 & 52.9 & 2.93 & 0.0143 & 1.33\\
SGMSE \cite{sgmse} & G & 30 & 770.2	& 273.6 & 1.55 & 0.0076 & 1.43 \\
\textbf{BinauralFlow (Ours)} & G & 6 & 163.0 &	314.5 & $\mathbf{1.00}$ & $\mathbf{0.0071}$ & $\mathbf{1.33}$\\
\bottomrule
\end{tabular}
}
\end{table*}

\begin{figure*}[t]
    \centering
    \includegraphics[width=\linewidth]{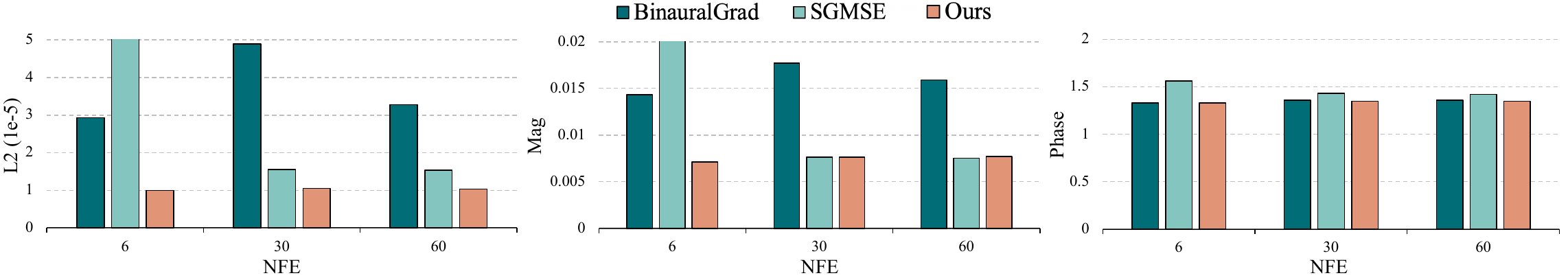}
    \vspace{-5mm}
    \caption{Performance with respect to the NFE. We evaluate all generative models using the same NFE for a fair comparison. }
    \label{fig:nfe}
\end{figure*}

\subsection{Experiment Details}
\noindent\textbf{Dataset.}
To evaluate BinauralFlow, we collect a new high-quality binaural dataset. We record 10 hours of paired mono and binaural data at $48$ kHz along with the head poses of the speaker and the listener. To match real-world scenarios, we collect data in a standard room without significant soundproofing or sound-absorbing materials. The background noise from multiple AC vents and electronic equipment is recorded. Furthermore, instead of using binaural mannequins and loudspeakers, both the speaker and the listener are real participants. During recording, the speaker is free to move anywhere in the room, and the listener is free to move the head while sitting on a chair. We split the dataset into training/validation/test subsets with 8.47/0.86/1.33 hours of each subset. The test subset contains two additional speakers, male and female, not seen during training. See the appendix for details on the data collection setup.

\noindent\textbf{Baselines.} We compare our approach with digital audio rendering approaches and more advanced neural audio rendering approaches. We choose SoundSpaces 2.0 \cite{chen2022soundspaces} as a DSP baseline given its powerful spatial audio rendering capability. For neural audio rendering models, we utilize 2.5D Visual Sound \cite{gao20192}, WaveNet \cite{van2016wavenet}, and WarpNet \cite{richard2021neural} as regression-based baselines, and use BinauralGrad \cite{leng2022binauralgrad} and SGMSE \cite{sgmse} as generative baselines. BinauralGrad is the state-of-the-art approach for the binaural speech synthesis task, which is a two-stage diffusion model.

\noindent\textbf{Metrics.} For quantitative evaluation, we leverage three metrics following WarpNet \cite{richard2021neural} and BinauralGrad \cite{leng2022binauralgrad}: waveform L2 error (L2), magnitude L2 error (Mag), and phase angular error (Phase).

\subsection{Quantitative Comparison}

\begin{figure}[t]
    \centering
    \includegraphics[width=\linewidth]{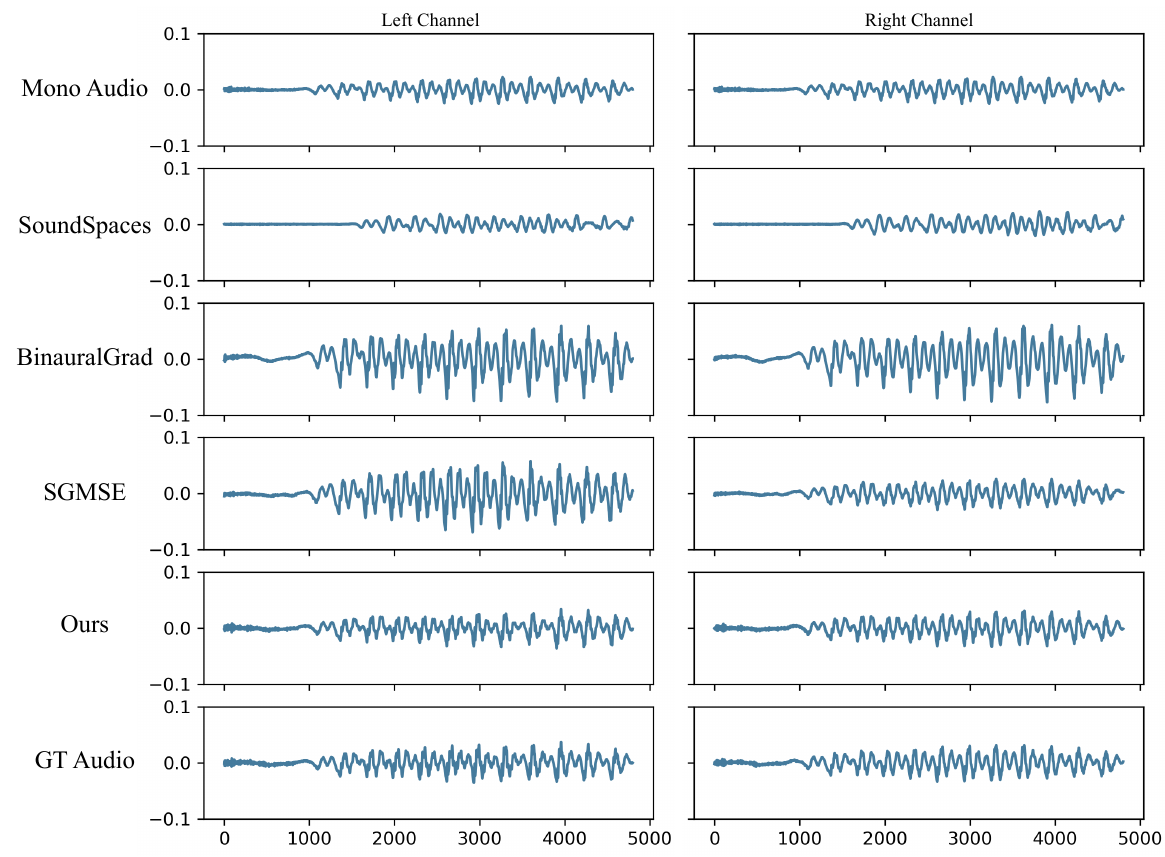}
    \vspace{-4mm}
    \caption{Qualitative comparison between different baselines. We display waveforms of rendered spatial audio.
    }
    \label{fig:visualization}
    \vspace{-4mm}
\end{figure}

We compare our method with existing baselines including the state-of-the-art approach BinauralGrad \cite{leng2022binauralgrad}. We present the metric results in \cref{tab:sota_speakeasy}, where lower values mean better performance. We show the number of function evaluations (NFE), i.e., how many times the model is called during binaural speech synthesis, and the model type for each method. As shown in the table, DSP and regression-based models underperform the generation-based models. Compared with BinauralGrad, SGMSE exhibits better generation quality in terms of L2 error and Mag error, but falls short in the Phase error. Our BinauralFlow model consistently outperforms all baselines with considerable margins. We also include the inference speed and the model size of each approach. We test the inference speed on a single 4090 GPU. The audio sampling rate is 48 kHz, and the audio length is 683 ms. Our model achieves the fastest inference speed among generative models. These results demonstrate that our model achieves a more favorable trade-off between performance and inference speed compared to the baseline approaches.

In \cref{tab:sota_speakeasy}, we use the default NFE of different generative models as recommended by the authors. We further evaluate these generative models using the same NFE for a more fair comparison. We test all models with $6,30,60$ NFEs and report the results in \cref{fig:nfe}, where each subfigure displays results for one metric.
As shown in the figure, our approach consistently outperforms other generative models across different NFEs, especially in the L2 and Mag metrics.

\begin{table*}[t]
    \centering
    \caption{Perceptual study. We report results of ABX, AB, and MUSHRA evaluation tasks, where higher values indicate better realism.} 
    \label{tab:perceptual}
    \resizebox{\linewidth}{!}{
    \begin{tabular}{l|c|cccc}
        \toprule
        Methods & NFE & ABX CR $\uparrow$ & A-B CR $\uparrow$ & Environment Score $\uparrow$ & Spatialization Score $\uparrow$\\
        \cmidrule(lr){3-4}
        \cmidrule(lr){5-6}
         &  & \multicolumn{2}{c}{{\tiny 0\% (clear difference wrt GT)} {\small--} {\tiny  50\% (random guess)}} & \multicolumn{2}{c}{{\tiny 0 (very different from reference/GT)} {\small--} {\tiny 100 (identical to reference/GT)}}\\
        \midrule
        SoundSpaces 2.0 \cite{chen2022soundspaces} & 1 & 5\% & 12\% & - & -\\
        BinauralGrad \cite{leng2022binauralgrad} & 6 & 4\% & 3\% & - & -\\
        SGMSE \cite{sgmse} & 30 & 11\% & 21\% & 41.1 $\pm$ 23.0 & 57.5$\pm$ 29.5 \\
        \textbf{BinauralFlow (Ours)} & 6 & $\mathbf{30\%}$ & $\mathbf{42\%}$ & $\mathbf{68.4\pm23.4}$ & $\mathbf{83.1\pm18.9}$\\
        \midrule
        Ground Truth & - & - & - & 87.4 $\pm$ 13.5 & 89.9 $\pm$ 11.1 \\
        \bottomrule
    \end{tabular}
    }
\end{table*}

\subsection{Qualitative Comparison}
To provide an intuitive comparison between different models, we display the waveforms of rendered binaural speech of various methods in \cref{fig:visualization}. The first row is the mono audio, the last row is the recorded audio, and the audios predicted by different methods are between them. The SoundSpaces approach estimates an inaccurate time delay between the transmitted mono audio and the received binaural audio. BinauralGrad and SGMSE predict accurate time delay but their amplitudes are mismatched. In comparison, our BinauralFlow model correctly predicts the time delay and audio amplitude. We show more results in the appendix. 

\subsection{Perceptual Study}
We conduct a comprehensive perceptual evaluation to assess the quality and realism of rendered outputs. When dealing with questions of the realism of generated samples, perceptual study is a more important indicator than numerical analysis because humans can perceive the authenticity of speech, and are sensitive to subtle but unnatural variations in sound, which is difficult to capture using purely numerical metrics. We perform the study in a quiet, acoustically treated room, with carefully calibrated playback levels and equalized headphones. See appendix for more details.

We recruit a total of $23$ participants and request them to complete the following tasks: 
\begin{itemize}[noitemsep,topsep=0pt,partopsep=0pt]
    \item ABX test: subjects are presented with 3 tracks, A, B, and X, and asked if X is A or if X is B (X is always one of them, and either A or B is the ground truth). This task measures if there is a perceivable difference between generated and recorded (ground truth) sounds.
    \item A-B test: subjects are presented with A and B and asked which they think is a real recording (one is always the ground truth). The task measures if users can reliably identify generated versus real sounds. 
    \item MUSHRA evaluation: subjects are presented with a reference (ground truth) and generated samples, and asked to rate their similarity in terms of environment (ambient noise and reverberation) and spatialization (sound source position). Scores range from 0 to 100, with higher scores indicating greater similarity.
\end{itemize} 
For the ABX and A-B tests, we define a confusion rate metric (CR) that calculates how often users confuse the rendered sound with the recorded one and make a wrong choice. The maximum value of a confusion rate is $50\%$, i.e. users cannot distinguish sounds and make random decisions. 

We show the perceptual evaluation results in \cref{tab:perceptual}. For all tasks, our approach outperforms other approaches with noticeable margins, showing remarkable rendering realism. In particular, in the A-B test we achieve a CR of $42\%$ (the upper bound is $50\%$), showing that users can barely distinguish our generated sounds from the recorded samples.

\subsection{Performance Analysis}
\label{sec:ablation}

We analyze the impacts of different design choices on our binaural speech synthesis framework.

\begin{table}[t]
\centering
\caption{Performance comparison between different flow matching approaches. The L2 error is on the scale of $1e{-}5$.}
\label{tab:flow_matching}
\resizebox{\linewidth}{!}{
\begin{tabular}{l|ccc}
\toprule
Methods & L2$\downarrow$ & Mag$\downarrow$ & Phase$\downarrow$ \\
\midrule
Simplified Flow Matching \cite{tong2023improving} & 1.86 & 0.0101 & 1.35\\
\textbf{BinauralFlow (Ours)} & $\mathbf{1.00}$ & $\mathbf{0.0071}$ & \textbf{1.33}\\
\bottomrule
\end{tabular}
}
\end{table}

\noindent\textbf{Flow Matching Methods.} In \cref{sec:flow}, we discuss the difference between proposed flow matching model and the simplified flow matching framework in \cite{tong2023improving}. Comparison results are shown in \cref{tab:flow_matching}. Our method achieves lower L2, Mag, and Phase errors, showing the effectiveness of our conditional flow matching approach.

\begin{figure}[t]
    \centering
    \includegraphics[width=0.9\linewidth]{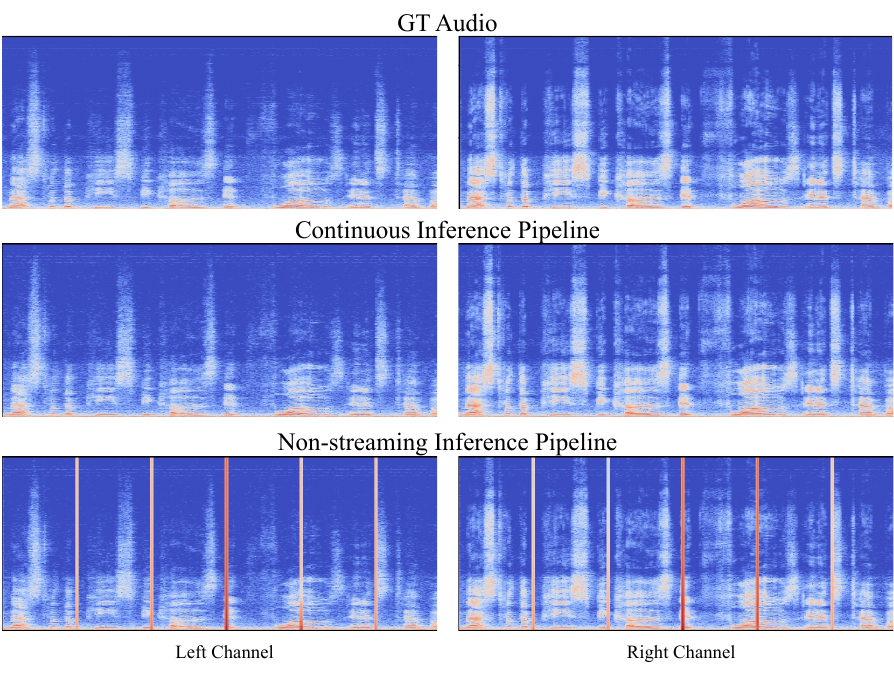}
    \vspace{-2mm}
    \caption{Output spectrograms using different inference pipelines.}
    \label{fig:streaming}
\end{figure}

\noindent\textbf{Continuous Inference Pipeline.} We compare our continuous inference pipeline and the non-streaming inference pipeline and show the generated spectrograms in \cref{fig:streaming}. Given a sequence of audio chunks, the non-streaming pipeline binauralizes each chunk individually, causing noticeable artifacts between adjacent chunks. In contrast, our pipeline synthesizes seamlessly smooth spectrograms.

\noindent\textbf{Real-Time Factor.} We calculate the real-time factor of our model for different numbers of function evaluations on a single 4090 GPU. The audio sampling rate is 48 kHz, and the audio length is 0.683 seconds. As shown in \cref{tab:rtf}, when NFE is set to 6, the real-time factor is 0.239. If we sacrifice some performance for faster inference, setting NFE to 1 results in an RTF of 0.04. Our model demonstrates potential for real-time streaming generation.
\begin{table}[t]
    \centering
    \caption{Real-time factor of our BinauralFlow model. We test the inference speed with different NFEs on a single 4090 GPU.}
    \begin{tabular}{c|cc}
    \toprule
        NFE	& Inference Time (sec) &	Real-Time Factor \\
    \midrule
        1 &	0.027 &	0.040 \\
        2 &	0.055 &	0.081 \\
        4 &	0.109 &	0.160 \\
        6 &	0.163 &	0.239 \\
        8 &	0.217 &	0.318 \\
        10 &	0.271 &	0.397 \\
    \bottomrule
    \end{tabular}
    \vspace{-3mm}

    \label{tab:rtf}
\end{table}

\begin{figure}[t]
    \centering
    \includegraphics[width=\linewidth]{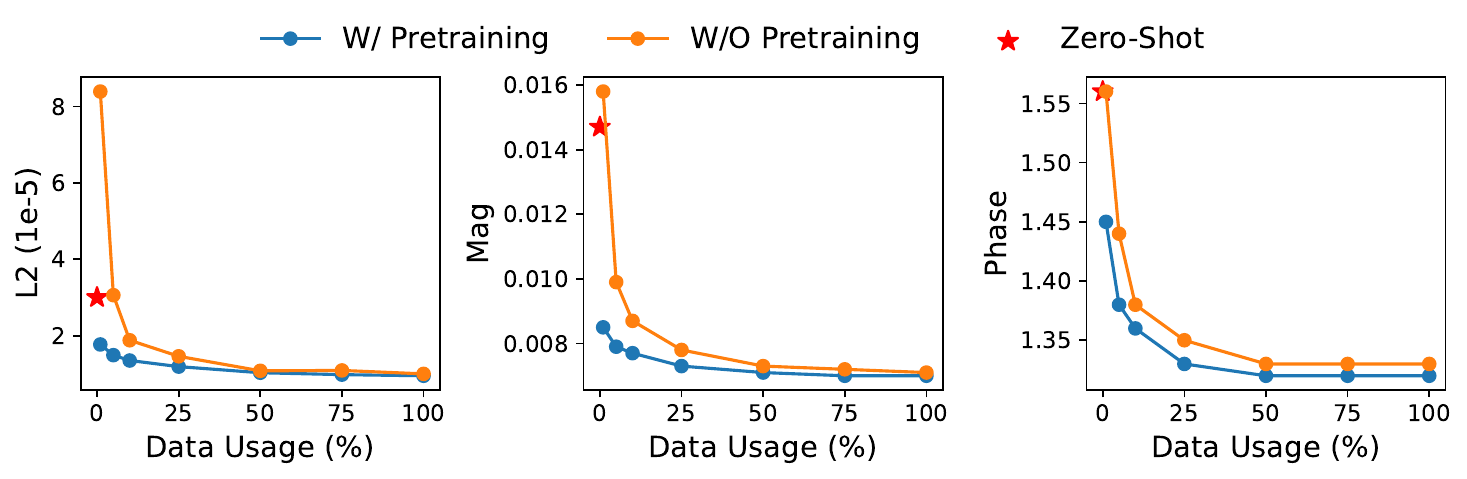}
    \vspace{-4mm}
    \caption{Large-scale pretraining strategy. We propose pretraining our model using massive data to improve data efficiency and enhance generalization in downstream tasks. }
    \vspace{-5mm}
    \label{fig:hearsay_pretrain}
\end{figure}

\noindent\textbf{Data Scale.} Recording 10 hours of data in real-world scenarios is costly and labor-intensive. To understand how data quantity affects our model's performance, we evaluate it using different amounts of training data ($1\%,5\%,10\%,25\%,50\%,75\%$). The results, shown in \cref{fig:hearsay_pretrain} (orange line), reveal a significant performance decline when using less than 25\% of the data.

To address this limitation, we develop a large-scale pre-training strategy using loudspeakers and artificial binaural heads instead of real individuals. While the use of artificial heads and loudspeakers reduces the quality and authenticity of the binaural data, it allows us to capture a large-scale dataset with over $7,700$ hours of binaural audio data, encompassing $97$ speaker identities from the English multi-speaker VCTK corpus \cite{Yamagishi2019VCTK} played by the loudspeaker. See the appendix for more details about the sytem and capture setup. 

We pretrain our BinauralFlow model on this dataset before fine-tuning it with limited real human data. As shown in \cref{fig:hearsay_pretrain} (blue lines), this pretraining strategy significantly improves performance. The pretrained model's zero-shot performance (red stars) matches or exceeds that of a model trained from scratch using only 1\% or 5\% real data. This demonstrates our model's robust generalization capabilities and its potential for various applications.
\section{Conclusion}
In this paper, we propose BinauralFlow, a streaming flow matching framework, that achieves high-quality continuous binaural speech rendering. Our framework consists of a conditional flow matching model, a causal U-Net architecture, and a continuous inference pipeline. Our framework surpasses existing baselines with significant improvement both quantitatively and qualitatively. A comprehensive perceptual study demonstrates that our model synthesizes binaural speech that is nearly indistinguishable from real recordings.

\newpage
\section*{Impact Statement}
Our work is designed to improve the rendering quality of binaural speech synthesis. 
It is not designed to modify the content of the input mono signal, but only to \textit{spatialize} it, i.e., place the source within an acoustic environment. 
However, we acknowledge that the enhanced realism may raise concerns about potential misuse, such as the creation of highly realistic deepfake audio. 
To address these risks, we emphasize the importance of adhering to ethical guidelines, fostering transparency in applications, and promoting responsible use of the proposed methods. Additionally, future research should focus on developing robust mechanisms for detecting and preventing misuse.


\bibliography{example_paper}
\bibliographystyle{icml2025}

%%%%%%%%%%%%%%%%%%%%%%%%%%%%%%%%%%%%%%%%%%%%%%%%%%%%%%%%%%%%%%%%%%%%%%%%%%%%%%%
%%%%%%%%%%%%%%%%%%%%%%%%%%%%%%%%%%%%%%%%%%%%%%%%%%%%%%%%%%%%%%%%%%%%%%%%%%%%%%%
% APPENDIX
%%%%%%%%%%%%%%%%%%%%%%%%%%%%%%%%%%%%%%%%%%%%%%%%%%%%%%%%%%%%%%%%%%%%%%%%%%%%%%%
%%%%%%%%%%%%%%%%%%%%%%%%%%%%%%%%%%%%%%%%%%%%%%%%%%%%%%%%%%%%%%%%%%%%%%%%%%%%%%%
\newpage
\appendix
\onecolumn
\newpage
\section{Demo Videos}
To help understand our work, we have created several demo videos showcasing BinauralFlow's binaural speech rendering capability. We include these demo videos on our webpage. We highly recommend that readers watch these videos to gain a deeper understanding of our research. In each video, we show a top-down view of the room along with the poses of the speaker and the listener. The speaker is denoted as ``Tx'' and the speaker's trajectory is shown in blue. The listener is denoted as ``Rx'' and the listener's trajectory is shown in red.

In the directory of each sample, we include two subdirectories: ``Comparison'' and ``Flip\_Test''. In the ``Comparison'' subdirectory, we display the results of SoundSpaces (``dsp''), BinauralGrad (``bgrad''), SGMSE (``sgmse''), and our BinauralFlow (``ours''). We also include the mono input (``mono'') and the recorded binaural audio (``gnd''). In the ``Flip\_Test'' subdirectory, we compare the synthesized sound and the ground-truth sound by using a flip-test technique. We periodically flip the sound between the synthesized sound and the ground-truth speech every $5$ seconds.

\section{Implementation Details}
We implement our streaming flow matching model with the PyTorch framework \cite{pytorch}. Our U-Net consists of seven Causal 2D Conv Blocks for the contracting and expanding parts. We only conduct the downsampling and upsampling operations four times. We set the window length as $512$, the hop length as $128$, and use a Hann window when applying STFT. The input audio length is $32768$ and the spectrogram is of shape $256\times 257$. We use the Adam optimizer \cite{adam} with a learning rate of $1e{-}4$ and a weight decay rate of $1e{-}5$. We set the standard deviation $\sigma$ of $\mathbf{z}$ as $0.5$. We use $6$ steps to solve the ODE with the midpoint solver and an early skip schedule.

\section{Midpoint Solver}
We present pseudo code of the inference process using the midpoint solver in \cref{alg:midpoint_solver}.

\begin{algorithm}[h]
\caption{Inference Procedure with Midpoint Solver}
\label{alg:midpoint_solver}
\begin{algorithmic}
    \STATE {\bfseries Input:} Trained network $u_{t}$, mono spectrogram $\mathbf{x}$, inference steps $n$
    \STATE $\mathbf{z} \sim \mathcal{N}(\mathbf{x},\sigma^2I)$ // \textcolor{gray}{{\scriptsize Sample random variable}}
    \STATE $t \gets 0$
    \STATE $\phi_t(\mathbf{z}) \gets \mathbf{z}$
    \STATE $\delta \gets 2/n$
    \WHILE{$t < 1$}
        \STATE $v' \gets u_t(\phi_t(\mathbf{z}), p_\mathrm{rx}, p_\mathrm{tx}, \mathbf{x};\theta)$ // \textcolor{gray}{{\scriptsize Calculate vector field at $t$}}
        \STATE $\phi_{t'}(\mathbf{z}) \gets \phi_t(\mathbf{z}) + v' \delta$ // \textcolor{gray}{{\scriptsize Calculate flow at $t+\delta$}}
        \STATE $v'' \gets u_{t'}(\phi_{t'}(\mathbf{z}), p_\mathrm{rx}, p_\mathrm{tx}, \mathbf{x};\theta)$ // \textcolor{gray}{{\scriptsize Calculate vector field at $t+\delta$}}
        \STATE $v=(v' + v'') / 2$ // \textcolor{gray}{{\scriptsize Average the vector field}}
        \STATE $\phi_t(\mathbf{z}) \gets \phi_t(\mathbf{z}) + v \delta$ // \textcolor{gray}{{\scriptsize Update the flow}}
        \STATE $t \gets t + \delta$ // \textcolor{gray}{{\scriptsize Update the time step}}
    \ENDWHILE
    \STATE $\mathbf{y} \gets \phi_t(\mathbf{z})$
    \STATE {\bfseries Output:} binaural spectrogram $\mathbf{y}$
\end{algorithmic}
\end{algorithm}
Given a trained network $u_t$, a mono spectrogram $\mathbf{x}$, and a predefined inference step $n$, we sample a random noise $\mathbf{z}$ following the Gaussian distribution $\mathcal{N}(\mathbf{x},\sigma^2I)$. We initialize some variables, including $t$, $\phi_t(\mathbf{z})$, and $\delta$. For each time step $t$, we calculate the vector field at two places, $t$ and $t+\delta$. Then we average these two vector fields and update the flow using the average vector field. In the end, we output updated $\phi_t(\mathbf{z})$ as the rendered binaural spectrogram $\mathbf{y}$.

\begin{figure}[t]
\centering
\subfigure[We collect data in a standard room without significant soundproofing or sound-absorbing materials.
The background noise from multiple air conditioning vents and electronic equipment is recorded.]{%
\label{fig:capture_setup}%
\includegraphics[width=0.45\linewidth]{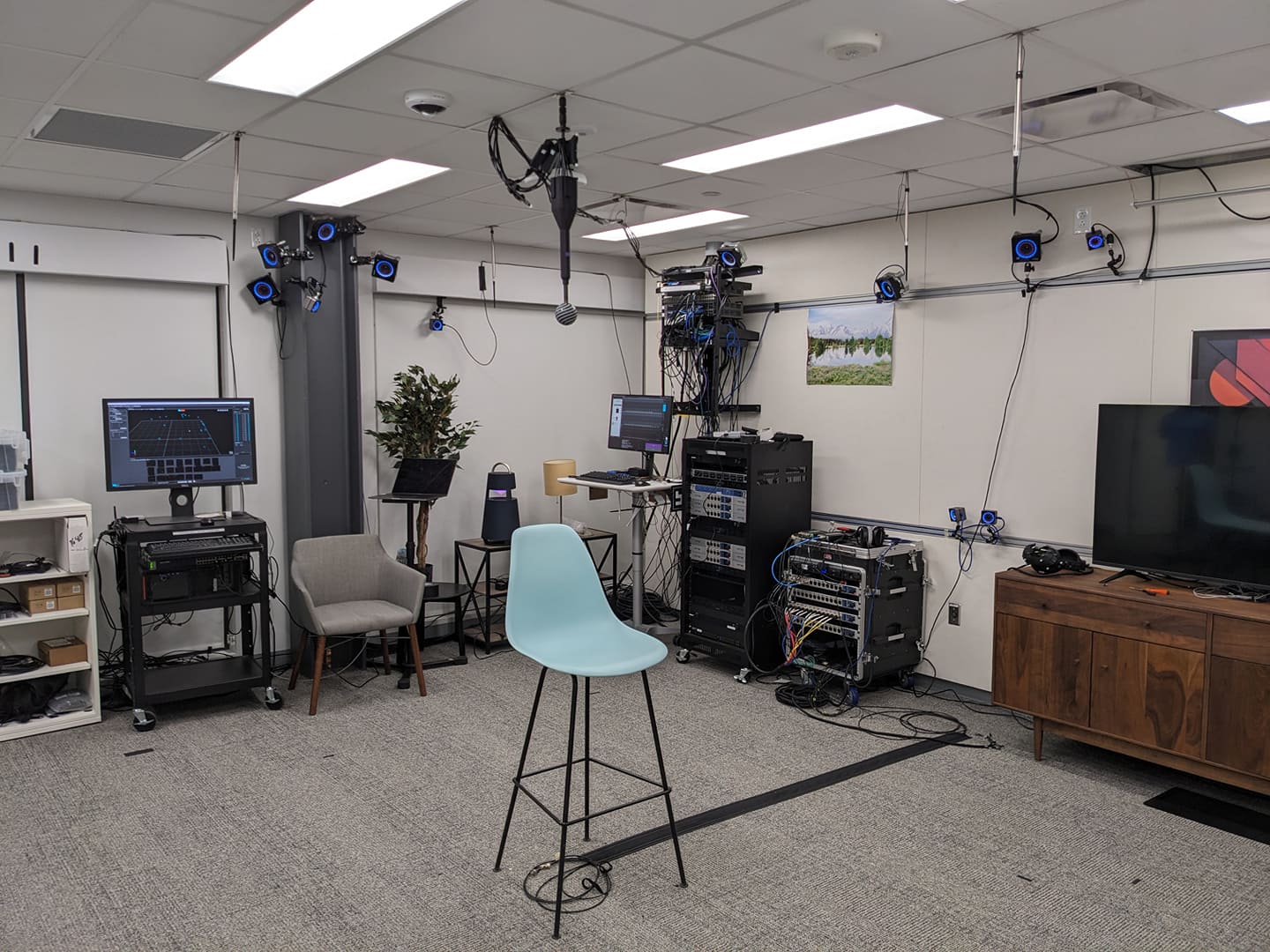}}%
\qquad
\subfigure[We perform the perceptual study in a quiet, acoustically treated room, with carefully calibrated playback levels and equalized headphones.]{%
\label{fig:evaluation_setup}%
\includegraphics[width=0.45\linewidth]{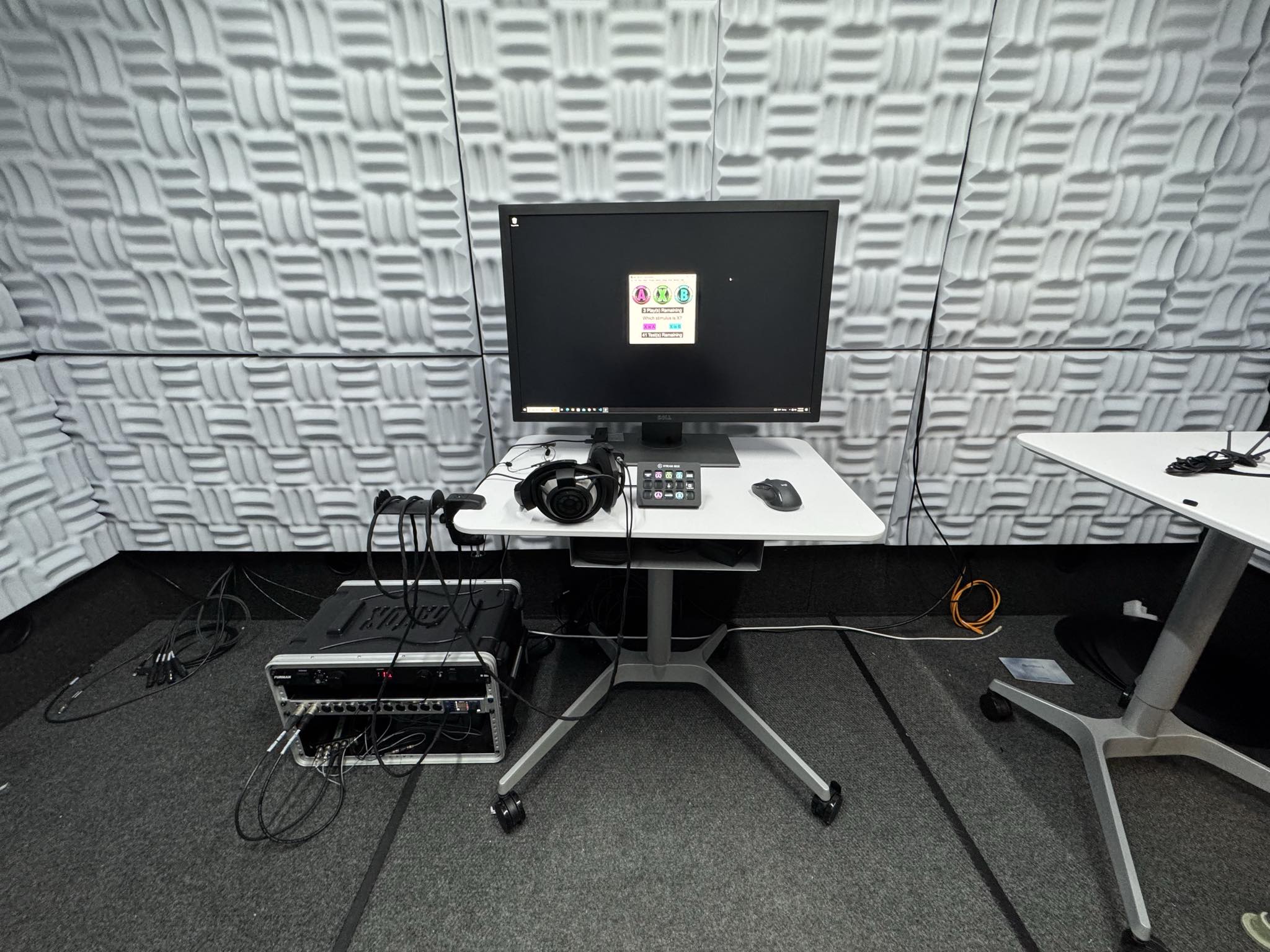}}%
\caption{Capture and evaluation setups.}
\end{figure}
\begin{figure}[h]
    \centering
    \includegraphics[width=0.5\linewidth]{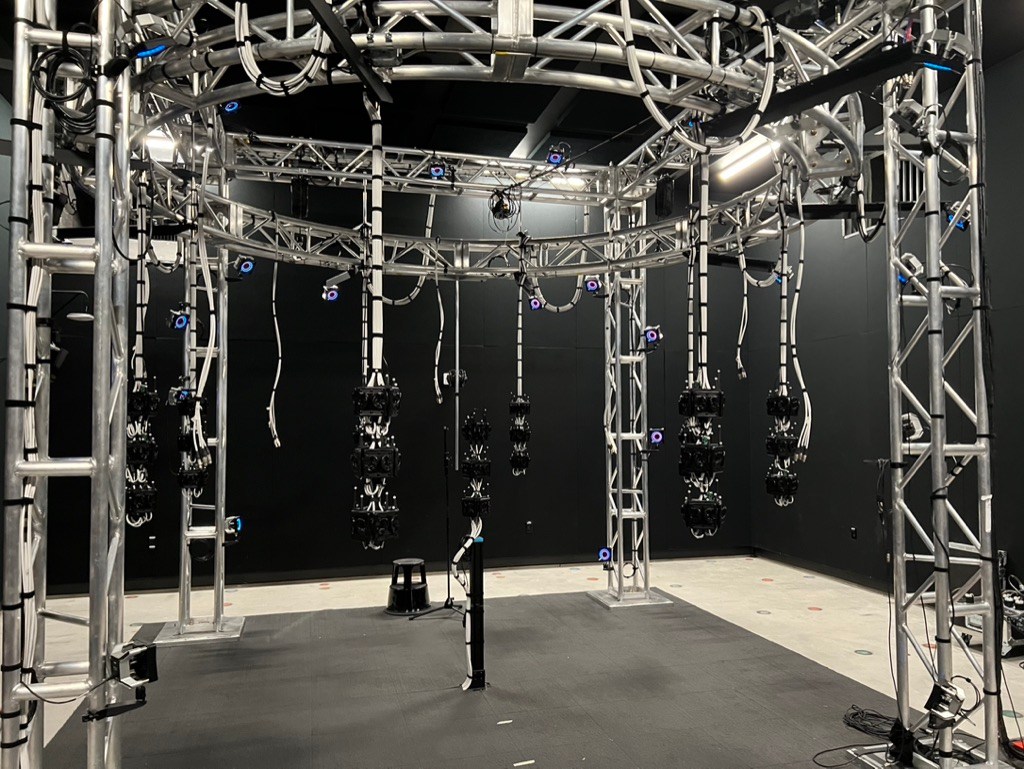}
    \caption{The large-scale binaural data capture system with artificial binaural heads.}
    \label{fig:hearsay}
\end{figure}

\section{Data Collection Setup}
The data collection featured a seated listener (they were free to move their head), and a speaker talking within approximately 1 m radius from the listener. 
A single participant acted as the listener, while three participants were captured as speakers, with one participant used as a part of the training set. The captures were performed in a non-anechoic room. The audio system featured a calibrated B+K 4101-B Binaural Microphone pair worn by the listener, as well as several DPA 4060s microphones mounted to a VR headset worn by the speaker, with guaranteed phase synchronization. Poses of both the speaker and listener were recorded via an OptiTrack tracking system. The speaker was tracked with IR-reflective markers mounted on the headset, and the listener was tracked via small facial IR-reflective markers. The setup is shown in \cref{fig:capture_setup}.

\section{Perceptual Study Setup}
The perceptual study's hardware setup included a PC workstation, RME 12mic + RME Digiface AVB, B\&K Type 4101 in-ear microphones, Sennheiser HD 800S headphones, and a Stream Deck. The study was conducted inside an 8' $\times$ 12' Whisper Room, with the monitor inside and the computer placed outside to ensure a controlled, noise-free environment. Custom Matlab and Max MSP patches were developed for use with in-ear mics to create a headphone equalization profile and recreate the recorded signal as accurately as possible. Participants were presented with a number of randomly chosen 4-second-long clips and had 10/10/5 minutes to complete the ABX/MUSHRA/AB sections of the evaluation using a Stream Deck and mouse for response input/selection. The setup is shown in \cref{fig:evaluation_setup}.

\section{Pre-training Dataset}
We set up 3Dio Omni binaural heads with human-shaped ears in a non-anechoic recording room. For data collection, operators walked around the room with a handheld loudspeaker, playing speech signals from the English multi-speaker VCTK corpus \cite{Yamagishi2019VCTK}. Our setup used 135 binaural heads and involved 33 loudspeaker operators. We used the OptiTrack system to track the 3D position and orientation of both the loudspeakers and the stationary binaural heads. We collected over $7,700$ hours of binaural audio data, encompassing $97$ speaker identities from the VCTK dataset across an area of $4.6$m horizontally and $2.4$m vertically. The audio was sampled at $48$ kHz, with tracking data recorded at $240$ frames per second. The setup is shown in \cref{fig:hearsay}.

\section{Comparison with Other Flow Matching-Based Audio Models}
PeriodWave \citep{lee2024periodwave} designs a multi-period flow matching model for high-fidelity waveform generation. FlowDec \citep{liu2024rfwave} introduces a conditional flow matching-based audio codec to noticeably reduce the postfiler DNN evaluations from 60 to 6. RFWave \citep{welker2025flowdec} proposes a multi-band rectified flow approach to reconstruct high-fidelity audio waveforms. These works are all related to flow matching models and show the effectiveness in generating high-quality waveform signals. To reduce the number of sampling steps, PeriodWave-Turbo \citep{lee2024accelerating} finetunes the CFM models with adversarial feedback. Matcha-TTS \citep{mehta2024matcha} employs a 1D U-Net model with 1D ResNet layers and Transformer Encoder layers. Neither the ResNet layers nor the Transformer Encoder layers are causal, which means that Matcha-TTS does not achieve time causality or support streaming inference. In contrast, our model is fully causal and supports streaming inference. CosyVoice 2 \citep{du2024cosyvoice} introduces a chunk-aware causal flow matching model that uses causal convolution layers and attention masks to enable causality. However, the CosyVoice 2 model does not include feature buffers for each causal convolution layer, which may result in audio interruptions and discontinuities during streaming inference in real-world scenarios.  

\section{Impact of Different Numerical Solvers}
Besides the Midpoint solver, we test the Euler and Heun solvers. The Euler solver is a first-order solver and Midpoint and Heun solvers are second-order. We set the number of function evaluations (NFE) to 6 and present the results in \Cref{tab:solver}. Although the Euler solver yields lower error values than the Midpoint solver, it fails to generate realistic background noise. Setting NFE to 6 is insufficient for the Heun solver, which requires 30 steps to achieve comparable error values. In conclusion, the Midpoint solver provides the best trade-off between error values, qualitative results, and inference efficiency.

\begin{table}[h]
    \centering
    \caption{Impact of different numerical solvers. We evaluate our model with various solvers, include both first-order and second-order solvers to analyze their influence on the generation quality.}
    \begin{tabular}{l|c|ccccc}
    \toprule
        Solver Type & NFE & Audio Quality & L2 $\downarrow$ & Mag $\downarrow$ & Phase $\downarrow$\\
        \midrule
        Euler &	6 &	Medium & 0.90 &	0.0066 &	1.24 \\
        Midpoint &	6 &	High &	1.00 &	0.0071 &	1.33 \\
        Heun	 & 6 &	Low	& 16.86	& 0.0499 &	1.44 \\
        Heun	 & 30 &	Medium &	1.27 &	0.0087 &	1.36 \\
    \bottomrule
    \end{tabular}

    \label{tab:solver}
\end{table}

\section{Sway Sampling Schedule}
\citet{chen2024f5} introduce a new timestep scheduler called Sway Sampling to improve inference quality and efficiency. We use Sway Sampling with different coefficients ranging from -1 to 1 to systematically evaluate its impact on our model. The results are shown in \cref{tab:sway}. Changing the coefficients does not lead to significant changes in the quantitative results. However, we observe that setting coefficients greater than 0, which shifts the time steps to the second half, results in better qualitative outcomes. Specifically, background noise becomes more realistic when the coefficient is increased. These results support the rationale behind our early skip strategy.

\begin{table}[h]
    \centering
    \caption{Impact of Sway Sampling with different coefficients.}
    \begin{tabular}{c|ccc}
    \toprule
        Coefficient & L2 $\downarrow$ & Mag $\downarrow$ & Phase $\downarrow$ \\
    \midrule
        -1.0 &	1.06 &	0.0070 &	1.29 \\
        -0.8 &	1.10 &	0.0070 &	1.29 \\
        -0.4 &	1.00 &	0.0069 &	1.29 \\
        0 &	1.02 &	0.0069 &	1.29 \\
        0.4 &	1.03	 & 0.0070 &	1.31 \\
        0.8	 & 1.04 &	0.0071 &	1.32 \\
        1.0 &	1.02 &	0.0072 &	1.33 \\
    \bottomrule
    \end{tabular}
    \label{tab:sway}
\end{table}

\section{Results on a Public Dataset}
In the main paper, we compare our BinauralFlow model with existing baselines on our own dataset. To further verify the effectiveness of our approach, we test our model on a public dataset released by \citet{richard2021neural}. We report the results in \cref{tab:sota_2hr}. As shown in the table, our model surpasses the state-of-the-art BinauralGrad in most of the metrics and performs on par with it in the Wave and Phase metrics.
\begin{table*}[h]
\small
\caption{Quantitative comparison with existing baselines on the public dataset. Wave L2 is on the scale of $\times 10^{-3}$.}
\label{tab:sota_2hr}
\begin{center}
\begin{tabular}{l|cccccc}
\toprule
Methods & PESQ $\uparrow$ & MRSTFT  $\downarrow$ & Wave L2 $\downarrow$ & Amplitude L2 $\downarrow$ & Phase L2 $\downarrow$ \\\midrule
DSP  & 1.610 & 2.750 & 1.543 & 0.097 & 1.596  \\
WaveNet \cite{van2016wavenet} & 2.305 & 1.915 & 0.179 & 0.037 & 0.968  \\
WarpNet \cite{richard2021neural} & 2.360 & 1.774 & 0.157 & 0.038 & 0.838  \\
BinauralGrad \cite{leng2022binauralgrad} & 2.759 & 1.278 & $\mathbf{0.128}$ & 0.030 & $\mathbf{0.837}$  \\
SGMSE \cite{sgmse} & 2.256 &	1.352	& 0.230	& 0.033	& 0.983 \\
\midrule
\textbf{BinauralFlow (Ours)} & $\mathbf{2.806}$ & $\mathbf{1.252}$ & 0.192 & $\mathbf{0.030}$ & 0.918 \\
\bottomrule
\end{tabular}
\end{center}
\end{table*}

\section{More Qualitative Results}
\begin{figure}[h]
    \centering
    \includegraphics[width=0.82\linewidth]{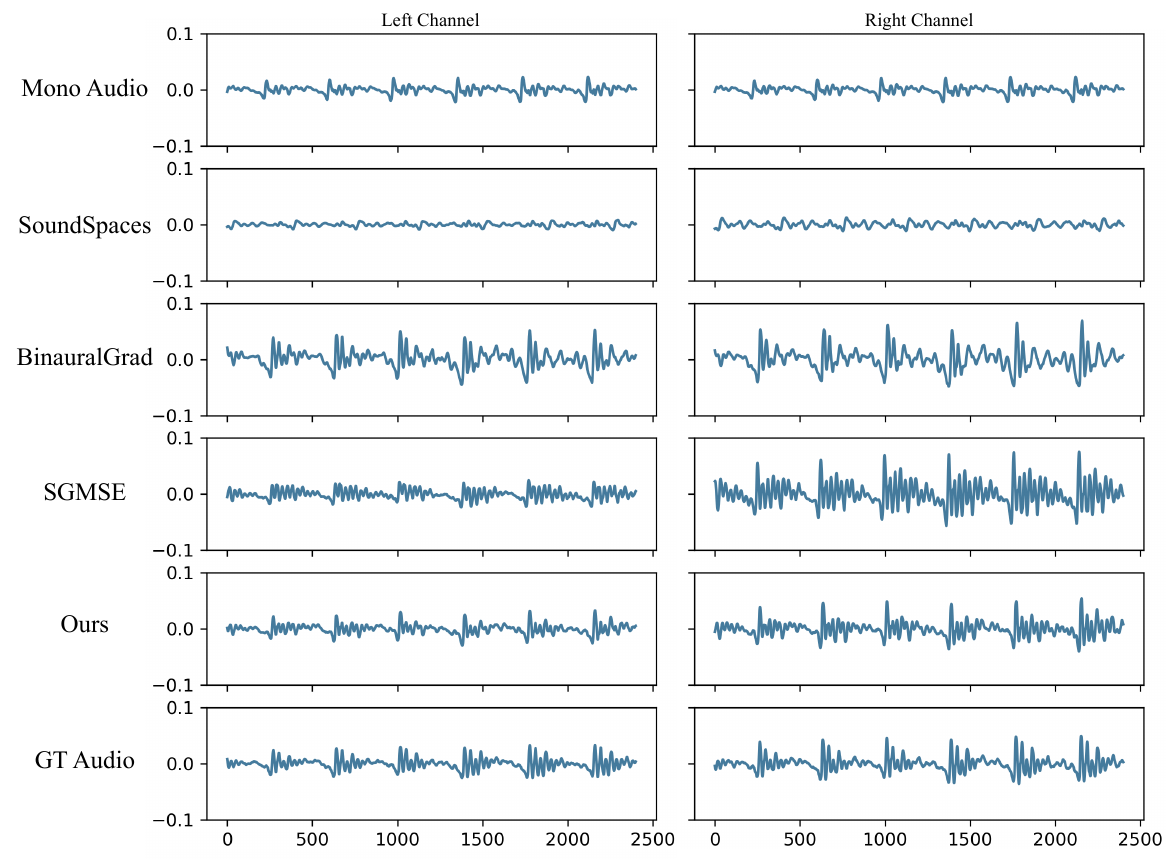}
    \caption{Qualitative comparison between different baselines. We display waveforms of rendered spatial audio.}
    \label{fig:visualization_apx1}
\end{figure}

\begin{figure}[h]
    \centering
    \includegraphics[width=0.82\linewidth]{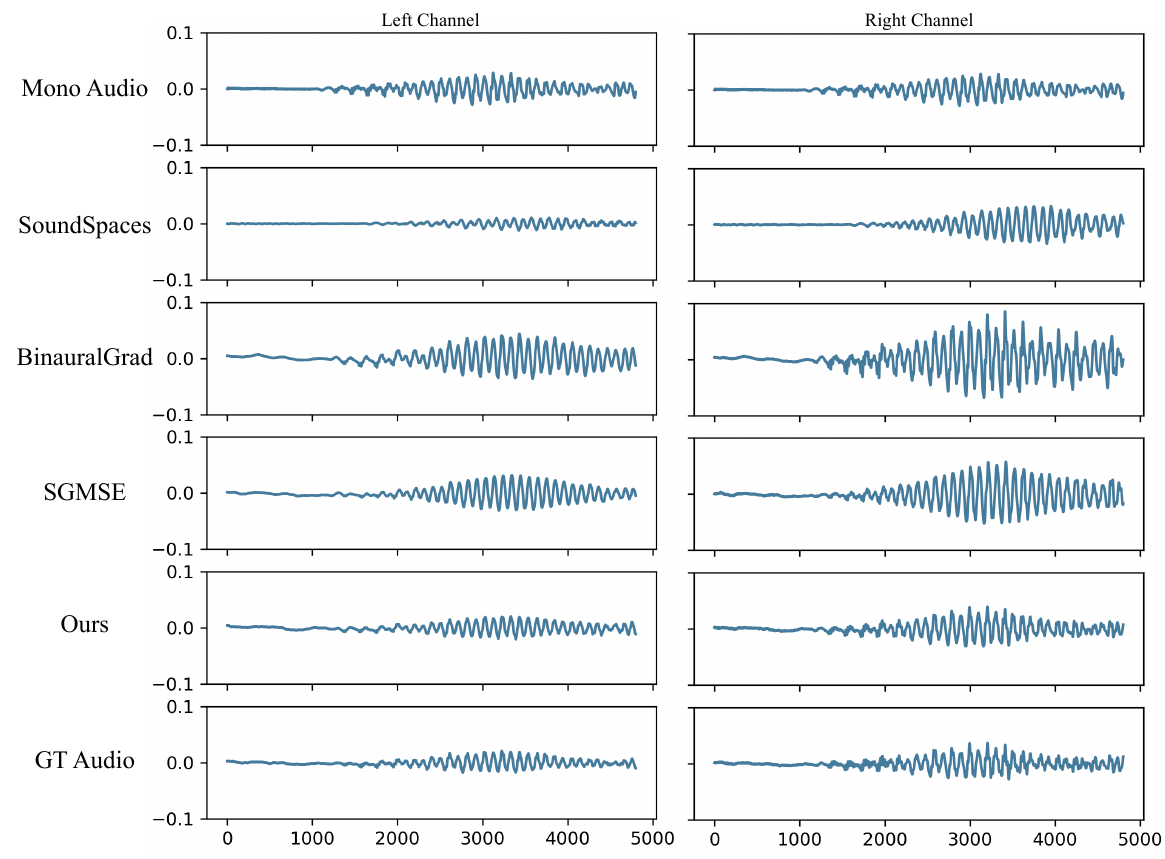}
    \caption{Qualitative comparison between different baselines. We display waveforms of rendered spatial audio.}
    \label{fig:visualization_apx2}
\end{figure}

\begin{figure}[h]
    \centering
    \includegraphics[width=0.82\linewidth]{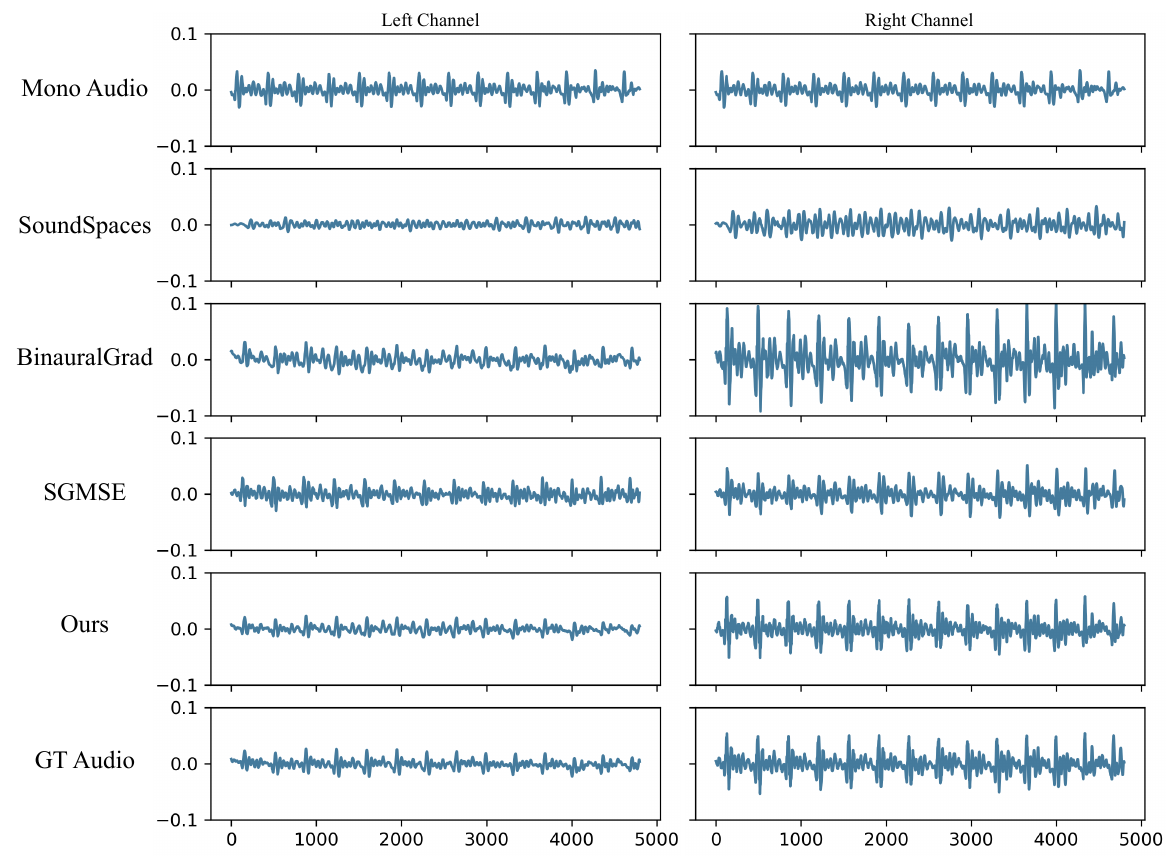}
    \caption{Qualitative comparison between different baselines. We display waveforms of rendered spatial audio.}
    \label{fig:visualization_apx3}
\end{figure}

We display more rendered waveforms in \cref{fig:visualization_apx1,fig:visualization_apx2,fig:visualization_apx3}. The first row is the mono audio, the last row is the recorded audio, and the audios predicted by different methods are between them. Our BinauralFlow model correctly predicts the time delay and audio amplitude.

\end{document}